\documentclass[aps,prc,reprint,superscriptaddress]{revtex4-1}
\usepackage[pdftex]{graphicx}
\usepackage[]{amsbsy}
\usepackage[]{amsmath}

\begin{document}


\title{Study of the astrophysically important $\boldsymbol{^{23}\mathrm{Na}(\alpha,p)^{26}\mathrm{Mg}}$ and $\boldsymbol{^{23}\mathrm{Na}(\alpha,n)^{26}\mathrm{Al}}$ reactions}


\author{M. L. Avila}
\email[]{mavila@anl.gov}
\affiliation{Physics Division, Argonne National Laboratory, Argonne IL 60439, USA}

\author{K. E.  Rehm}
\affiliation{Physics Division, Argonne National Laboratory, Argonne IL 60439, USA}

\author{S. Almaraz-Calderon}
\affiliation{Department of Physics, Florida State University, Tallahassee, FL 32306, USA}

\author{A. D. Ayangeakaa}
\affiliation{Physics Division, Argonne National Laboratory, Argonne IL 60439, USA}

\author{C. Dickerson}
\affiliation{Physics Division, Argonne National Laboratory, Argonne IL 60439, USA}

\author{C. R. Hoffman}
\affiliation{Physics Division, Argonne National Laboratory, Argonne IL 60439, USA}
 
\author{C. L. Jiang}
\affiliation{Physics Division, Argonne National Laboratory, Argonne IL 60439, USA}

\author{B. P. Kay}
\affiliation{Physics Division, Argonne National Laboratory, Argonne IL 60439, USA}

\author{J. Lai}
\affiliation{Department of Physics and Astronomy, Louisiana State University, Baton Rouge, LA, 70803, USA }

\author{O. Nusair}
\affiliation{Physics Division, Argonne National Laboratory, Argonne IL 60439, USA}

\author{R. C. Pardo}
\affiliation{Physics Division, Argonne National Laboratory, Argonne IL 60439, USA}

\author{D. Santiago-Gonzalez}
\affiliation{Department of Physics and Astronomy, Louisiana State University, Baton Rouge, LA, 70803, USA }
\affiliation{Physics Division, Argonne National Laboratory, Argonne IL 60439, USA}

\author{R. Talwar}
\affiliation{Physics Division, Argonne National Laboratory, Argonne IL 60439, USA}

\author{C. Ugalde}
\affiliation{Physics Division, Argonne National Laboratory, Argonne IL 60439, USA}


\date{\today}

\begin{abstract}
The $^{23}$Na$(\alpha,p)^{26}$Mg and $^{23}$Na$(\alpha,n)^{26}$Al reactions are important for our understanding of the $^{26}$Al abundance in massive stars. The aim of this work is to report on a direct and simultaneous measurement of these astrophysically important reactions using an active target system. The reactions were investigated in inverse kinematics using $^{4}$He as the active target gas in the detector. We measured the excitation functions in the energy range of about 2 to 6 MeV in the center of mass. We have found that the cross sections of the $^{23}$Na$(\alpha,p)^{26}$Mg and the $^{23}$Na$(\alpha,n)^{26}$Al reactions are in good agreement with previous experiments, and with statistical model calculations. 

\end{abstract}


\maketitle

\section{Introduction}

The importance of the radioisotope $^{26}$Al in the fields of $\gamma$-ray astronomy and chemical cosmology has been well established over the past few years.  The $\gamma$-rays from the decay of $^{26}$Al are direct evidence for the continuing nucleosynthesis in stars providing a unique way of testing the predictive power of theoretical stellar models. Since the half-life of $^{26}$Al ($7.2\times10^{5}$ years) is small compared to the time scales of Galactic chemical evolution $(\approx$ 10$^{10}$ years), the $^{26}$Al found in the interstellar medium is the outcome of relatively recent nucleosynthesis in the Galaxy.  $^{26}$Al can be traced by measuring the 1.809 MeV $\gamma$-ray line associated with its radioactive decay. 
A complete sky map of the corresponding $\gamma$-ray emission has been produced using data gathered by instruments on board the COMPTEL \cite{Diehl95} and INTEGRAL \cite{Diehl06} satellites. Although the precise source of $^{26}$Al is not completely understood, the observations by COMPTEL and INTEGRAL showed that the $^{26}$Al distribution is confined towards the Galactic disk, which strongly suggest massive stars ($M >8 M_{\odot}$) as one of the most likely production sites.

Sensitivity studies performed by \citet{Illiadis11} used around 900 nuclear reaction network calculations to determine the nuclear reactions that affect the $^{26}$Al abundance in massive stars. In their work, three different massive star sites were investigated: explosive neon-carbon burning, convective shell carbon burning and convective core hydrogen burning. The $^{23}$Na($\alpha,p)^{26}$Mg reaction, along with four other reactions, were suggested as prime targets for experimental measurements. The $^{23}$Na($\alpha,p)^{26}$Mg reaction is an important proton source for producing $^{26}$Al from $^{25}$Mg. Improved experimentally determined reaction rates for the $^{23}$Na($\alpha,p)^{26}$Mg reaction were needed at about 2.3 GK for explosive Ne/C burning and 1.4 GK for convective shell C/Ne burning.  Although the $^{23}$Na($\alpha,p)^{26}$Mg reaction was measured in Refs. \cite{Kuperus64,Whitmire74}, the authors of Ref. \cite{Illiadis11} did not considered the data reliable, due to problems associated with degradation of the NaCl targets used in the experiments, and the reaction rates used in their calculations theoretically estimated.

In Ref. \cite{Almaraz14}, the $^{23}$Na($\alpha,p)^{26}$Mg reaction was measured in inverse kinematics using $^{23}$Na beams of different energies impinging on a cryogenic $^4$He gas target. This study reported a reaction rate which was higher than the recommended rate by about a factor of 40. This result would have significant implications for the $^{26}$Al production, suggesting that the abundance of $^{26}$Al was larger by a factor of about 3. 
More recently, this reaction was re-measured \cite{Tomlinson15,Howard15}. The experiment reported in Ref. \cite{Tomlinson15} used inverse kinematics with a $^{23}$Na beam impinging on a $^4$He gas target, similar to the experiment of Ref. \cite{Almaraz14}. The measurement of Ref. \cite{Howard15} was performed in normal kinematics with a $^4$He beam of energies between 1.99 and 2.94 MeV bombarding a carbon-backed NaCl target. The measurements of Refs. \cite{Tomlinson15,Howard15} were found to be in good agreement with each other and with statistical model calculations, however they were in disagreement with the large cross sections found in Ref. \cite{Almaraz14}. Due to these discrepancies, the data of Ref. \cite{Almaraz14} was reanalyzed and an error in the normalization was found \cite{Almaraz15}, which overestimated the cross section by a factor of 100. The corrected cross section are in agreement with the results reported in Ref. \cite{Tomlinson15,Howard15}. The goal of the present work was to repeat the measurement with an independent technique, using an active target and detector system that measures both the $^{26}$Mg recoils from the $^{23}$Na$(\alpha,p)^{26}$Mg reaction and the incoming $^{23}$Na beam with the same detector, removing the need for a normalization common to all previous measurements.

Another important process in nuclear astrophysics is the $^{26}$Al($n,\alpha)^{23}$Na reaction which is one of the dominant destruction mechanisms of $^{26}$Al. The sensitivity studies performed in Ref. \cite{Illiadis11} have reported a strong dependence of the $^{26}$Al yield on this reaction. The experimental efforts measuring this reaction, have mainly focused on the time-inverse $^{23}$Na($\alpha,n)^{26}$Al reaction. This is primarily because of difficulties associated with the fabrication of a radioactive $^{26}$Al target. Therefore, the study of the $^{23}$Na($\alpha,n)^{26}$Al reaction contributes to our understanding of the $^{26}$Al abundance. Refs. \cite{Norman82} and \cite{Skelton87} reported on the study of the $^{23}$Na($\alpha,n)^{26}$Al reaction and applied the principle of detailed balance to obtain the contribution of the $^{23}$Na($\alpha,n_0)^{26}$Al reaction. A disadvantage of studying the time-inverse reaction is that it only provides information of the ground state transition of $^{23}$Na and contributions from excited states have to be calculated using theoretical models. However, the study of this reaction can be used to apply constrains to the $^{23}$Na($\alpha,n)^{26}$Al reaction.

In the present work the $^{23}$Na($\alpha,p)^{26}$Mg and $^{23}$Na($\alpha,n)^{26}$Al reactions are measured simultaneously. Thus, problems related to different detection systems, efficiencies and normalization of the cross sections are avoided.

\section{Experimental setup}

The experiment was carried out at the ATLAS accelerator at Argonne National Laboratory. The data have been measured using a Multi-Sampling ionization Chamber (MUSIC) detector with a close to 100\% detection efficiency. This detector has been previously used for measurements of fusion reactions of astrophysical interest \cite{Carnelli15}. MUSIC is an active target system with 18 anode strips allowing the measurement of an excitation function covering a large energy range. A full description of the detector and a more detailed explanation of the operation principles can be found in Ref. \cite{Carnelli15}. The technique for the measurement of $\alpha$-particle induced reactions has been benchmarked with the $^{17}$O$(\alpha,n)^{20}$Ne reaction for which cross sections can be found in the literature \cite{Bair73}. More details about the data analysis of $(\alpha,p)$ and $(\alpha,n)$ reactions (including the $^{23}$Na$+^4$He system discussed in the present work) as well as verification of the technique will be published in a separate paper \cite{AvilaNIM16}. 

The experiment was performed in inverse kinematics using $^{23}$Na beams with energies of 51.5 and 57.4 MeV, and intensities up to 5000 particles/sec.  To reduce the beam intensity, a series of pepper-pot attenuators \cite{Kubik87} and the ATLAS beam sweeper, which increased the pulse period of the beam from 82 ns to 41 $\mu$s, were used. The beam was delivered to the MUSIC detector which was filled with 403 and 395 Torr of $^4$He gas for the lower and higher energy measurement, respectively. With these combinations of energies and pressures, an energy range in the center of mass of $E_{c.m.}=2.2$-5.8 MeV was covered. In this energy range, both $(\alpha,p)$ ($Q=1.820$ MeV) and $(\alpha,n)$ ($Q=-2.967$ MeV) channels are open, allowing us to measure the $^{23}$Na($\alpha,p)^{26}$Mg and $^{23}$Na($\alpha,n)^{26}$Al reactions simultaneously.

\begin{figure}
\centering
\includegraphics[scale=0.43]{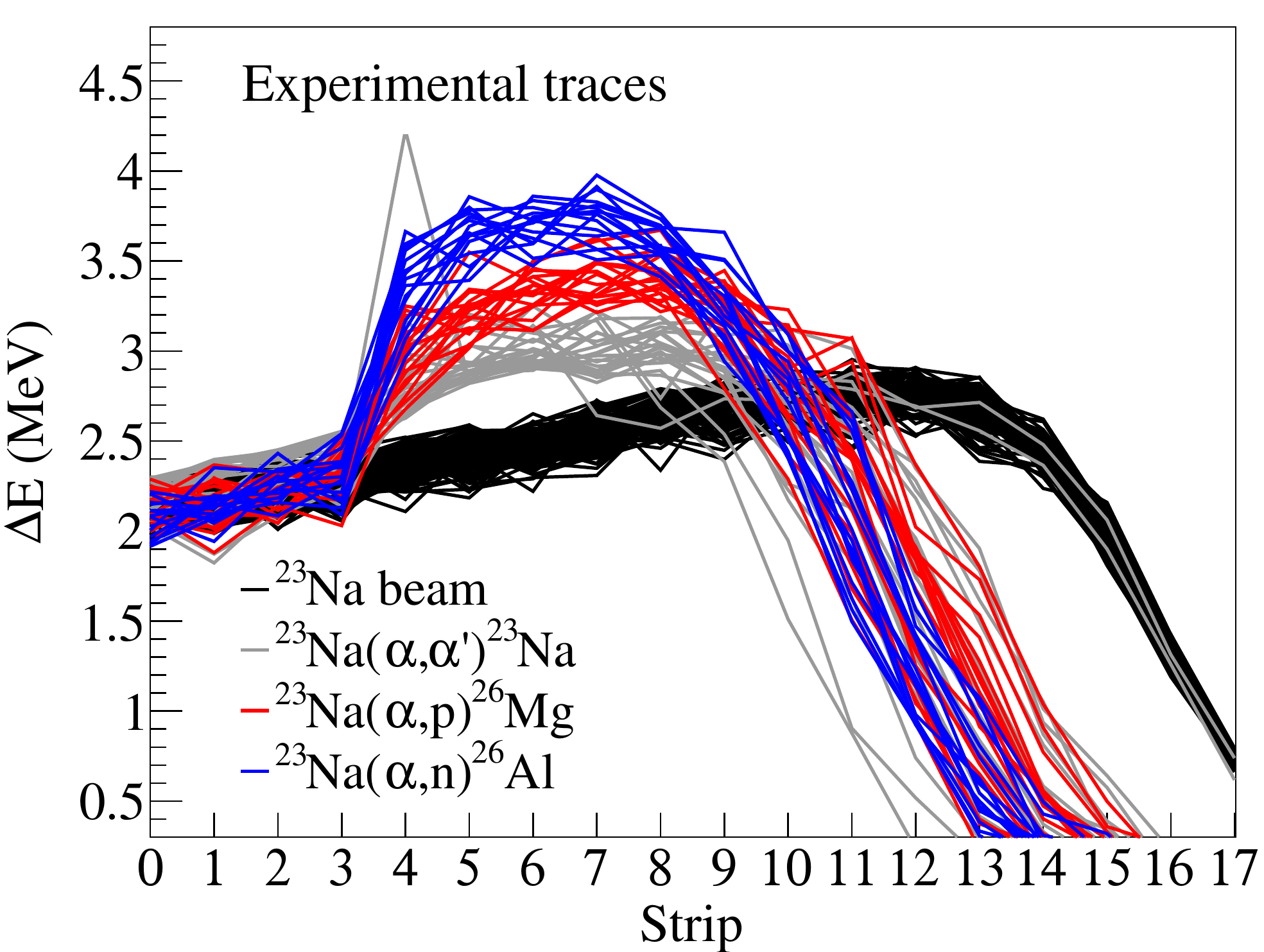} 
\caption{\label{fig:traces_23Na} 
(Color online) Energy-loss signals measured in the 18 strips of the MUSIC detector for events of the $^{23}$Na$(\alpha,n)^{26}$Al (blue), $^{23}$Na$(\alpha,p)^{26}$Mg (red) and $^{23}$Na$(\alpha,\alpha')^{23}$Na (gray) reactions occurring in strip 4. The black lines originate from the $^{23}$Na beam. }
\end{figure}

The separation of events from the two reactions is performed by analyzing the differences in the energy loss ($\Delta$E) of the reaction products in each strip of the detector. For example, the $\Delta$E for reactions occurring in strip 4 for a one-hour run from the higher beam energy measurement can be seen in Fig. \ref{fig:traces_23Na}. In this figure, four groups of traces with different $\Delta$E values, which originate from the $^{23}$Na beam (black), the $(\alpha,p)$ reaction (red), the $(\alpha,n)$ reaction (blue) and from elastic and inelastic scattering reaction (gray), are visible. For a better visualization only the first 25 events of the $(\alpha,\alpha')$ reaction are shown. The energy of the $^{23}$Na ions passing through strip 0 was about 39 MeV and for a pressure of 395 Torr the beam was almost stopped at strip 16, as can be seen in Fig. \ref{fig:traces_23Na}. The experimental traces seen in  Fig. \ref{fig:traces_23Na} are in agreement with simulated traces \cite{AvilaNIM16}.

\begin{figure}
\centering
\includegraphics[scale=0.44]{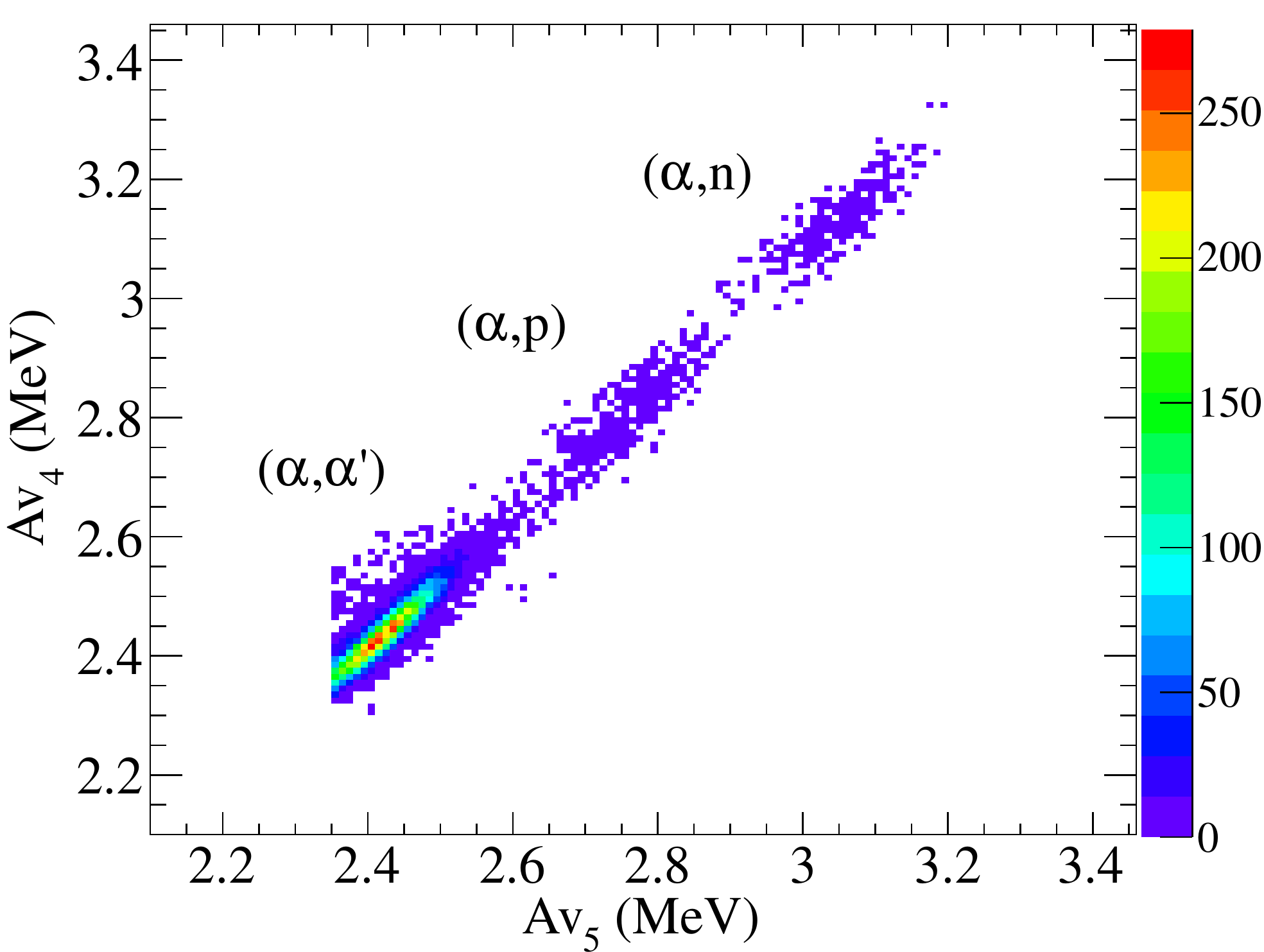} 
\caption{ \label{fig:Ave1_Ave2} 
(Color online) Two-dimensional plot of $\Delta$E values for events occurring in strip 4, averaged over five (Av$_5$) and four strips (Av$_4$) in order to improve the separation of events from the $^{23}$Na$(\alpha,\alpha')^{23}$Na, the $^{23}$Na$(\alpha,p)^{26}$Mg and the $^{23}$Na$(\alpha,n)^{26}$Al reactions.. }
\end{figure}

To improve the separation of the three reactions we have averaged the $\Delta$E values over a certain number of strips following the strip where the reaction took place, as explained in Ref. \cite{AvilaNIM16}. The average is called Av$_n$ with $n$ indicating the number of strips used to calculate the average. In Fig. \ref{fig:Ave1_Ave2}, a two-dimensional plot of a five-strip average (Av$_5$) against a four-strip average (Av$_4$) for the whole 1.5 days long run is shown. The sharp cut seen in this figure at 2.3 MeV in the $x$-axis (Av$_5$) is due to a condition applied to the data in order to discard the beam-like events. In this figure, the three groups originating from $(\alpha,\alpha')$, $(\alpha,p)$ and $(\alpha,n)$ reactions are clearly distinguishable.

With this approach the cross sections of the $^{23}$Na$(\alpha,p)^{26}$Mg and $^{23}$Na$(\alpha,n)^{26}$Al reactions have been determined covering the energy range E$_{c.m.}\approx$2-6 MeV in the center-of-mass frame. The normalization of the cross section is performed by using the number of beam particles (black traces in Fig. \ref{fig:traces_23Na}) which are simultaneously measured in the detector.

\section{Results}

\begin{figure}
\centering
\includegraphics[scale=0.3]{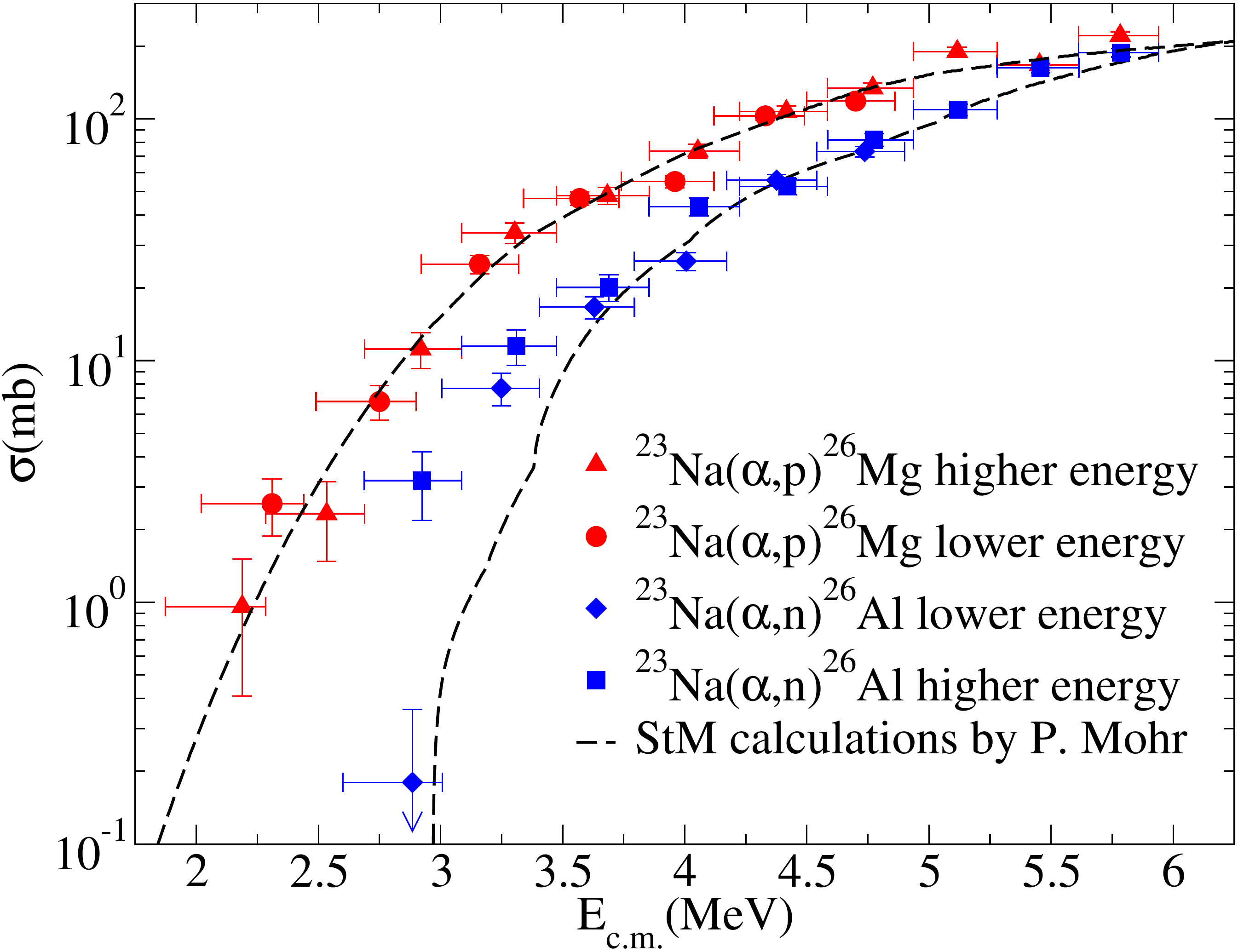} 
\caption{\label{fig:ap_an_comparison} 
(Color online) Excitation functions  of the $^{23}$Na($\alpha,n)^{26}$Al (blue) and the $^{23}$Na($\alpha,p)^{26}$Mg (red) reactions measured with  the MUSIC detector for a beam energies of 51.5 and 57.4 MeV labeled as low energy and high energy, respectively.}
\end{figure}

The excitation functions of angle- and excitation-energy- integrated cross sections of the $^{23}$Na($\alpha,p)^{26}$Mg and $^{23}$Na($\alpha,n)^{26}$Al reactions were measured in two runs lasting about 1.5 days each for the higher and lower energy, respectively. The results are presented in Fig. \ref{fig:ap_an_comparison}, where the ($\alpha,p)$ data are shown by red circles for the lower beam energy and by red triangles for the higher beam energy. Similarly, the ($\alpha,n)$ cross sections are shown by blue diamonds for the lower energy and blue squares for the higher beam energy, respectively. The uncertainties in the cross sections are statistical and the uncertainties in the center of mass energy are due to the energy range in each anode strip. The dashed lines are the cross sections predicted for the two reactions by the statistical model from Ref. \cite{Mohr15} using the TALYS code. The energy in the middle of each strip was calculated using the energy loss values of the code SRIM (version 2008) \cite{SRIM}. If the energy loss values are taken from the LISE++ prediction \cite{LISE}, there is an energy difference of about 10\% on average. For the data points seen in Fig. \ref{fig:ap_an_comparison}, an effective energy was calculated instead of using the energy in the middle of each strip in order to take into account the energy-dependence of the cross section. The energy width of a given strip average the cross section over $\sim$320 keV for the first strips and $\sim$500 keV for the last strips in the center-of-mass system.

\subsection{The $\boldsymbol{^{23}\mathrm{Na}(\alpha,p)^{26}\mathrm{Mg}}$ reaction}

\begin{figure}
\centering
\includegraphics[scale=0.3]{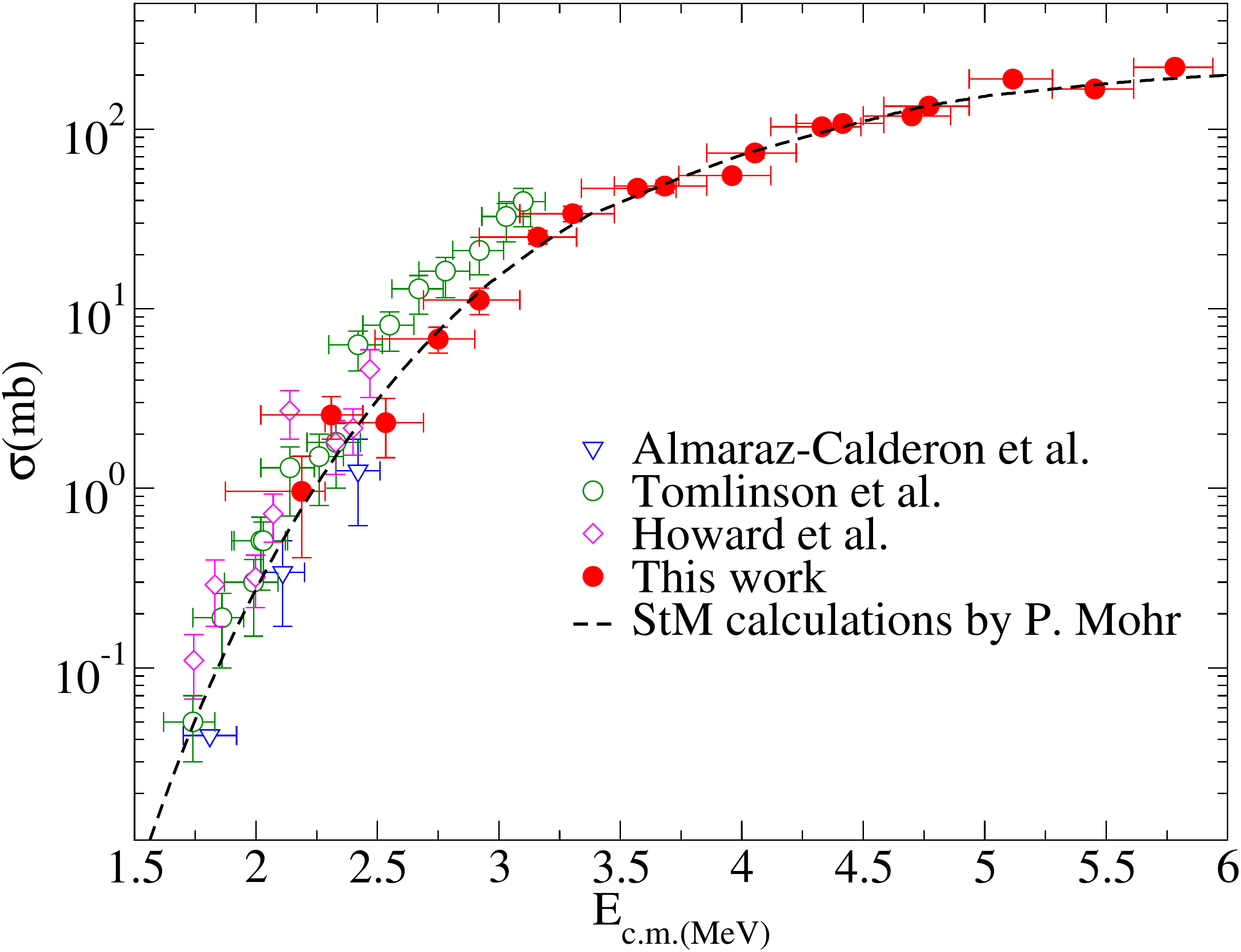} 
\caption{\label{fig:ap_total} 
(Color online) Excitation functions of the $^{23}$Na($\alpha,p)^{26}$Mg reaction obtained in this work (red solid circles), in comparison with the results from \citet{Almaraz15} (blue triangles), \citet{Tomlinson15} (green open circles), \citet{Howard15} (magenta diamonds) and the statistical model calculations from \citet{Mohr15}.}
\end{figure}

The $^{23}$Na$(\alpha,p)^{26}$Mg reaction was previously studied in the energy range of about 1.7-3 MeV in the center-of-mass system. Our experiment overlaps with previous measurements in the energy range of about 2-3 MeV. The cross section obtained in this work is found to be in good agreement with the experiments of Refs. \cite{Tomlinson15,Howard15,Almaraz15} and with the statistical model calculations of Ref. \cite{Mohr15}. Moreover, our experiment was able to extend the measurements towards higher energies (up to 6 MeV) where no experimental data existed. In this energy region we again notice that the cross sections measured are in agreement with the statistical model calculations from Ref. \cite{Mohr15}. A comparison of the $^{23}$Na($\alpha,p)^{26}$Mg cross sections from this experiment with previous measurements is presented in Fig. \ref{fig:ap_total}. 

Our work confirms the cross sections and the reaction rates obtained in previous experiments and from statistical models calculations. 

\subsection{The $\boldsymbol{^{23}\mathrm{Na}(\alpha,n)^{26}\mathrm{Al}}$ reaction}

\begin{figure}
\centering
\includegraphics[scale=0.3]{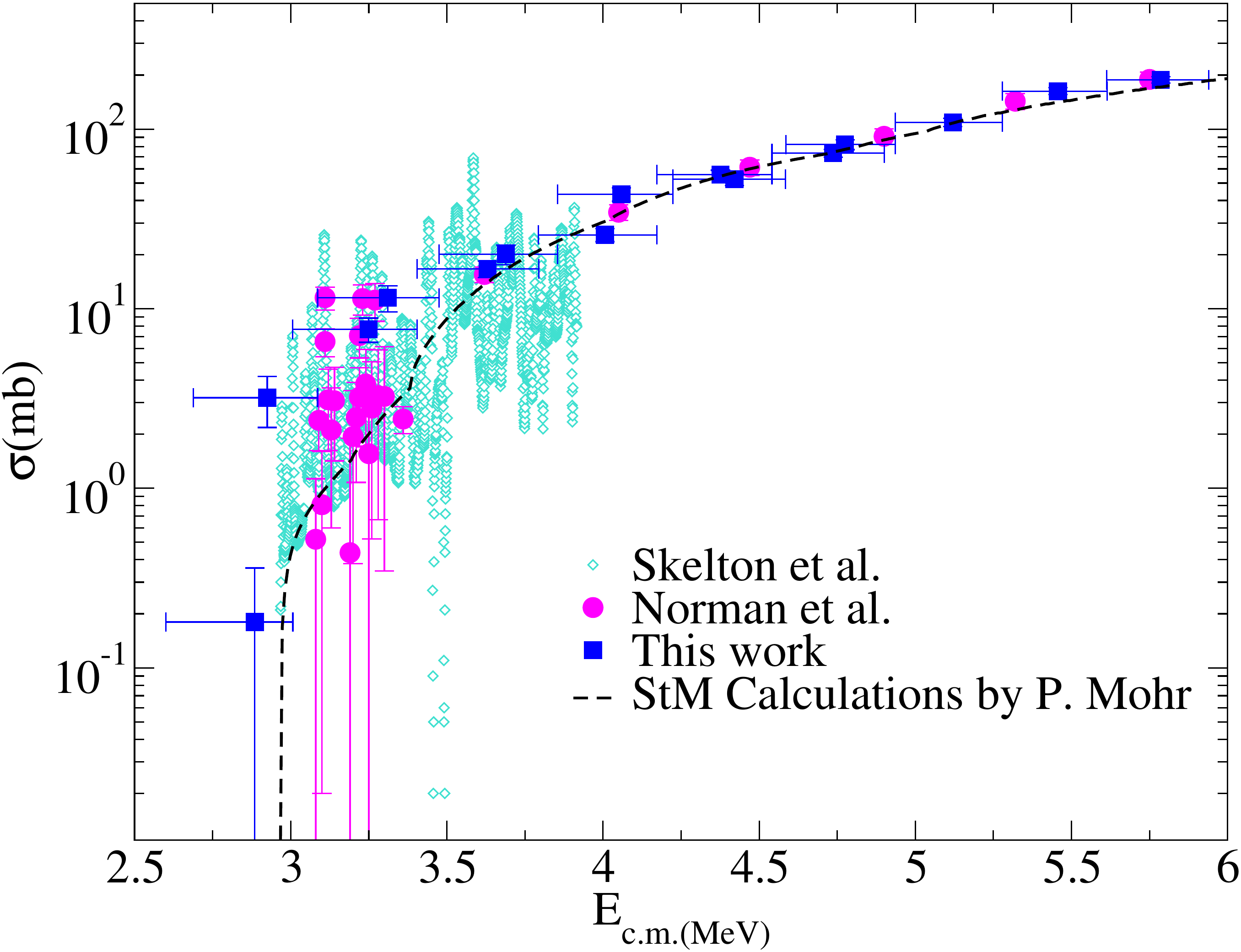} 
\caption{\label{fig:an_total} 
(Color online) Excitation functions of the $^{23}$Na($\alpha,n)^{26}$Al reaction obtained in this work (blue squares), by \citet{Norman82} (magenta circles), \citet{Skelton87} (turquoise diamonds), and from \citet{Mohr15}.}
\end{figure}

In our experiment we have measured the $^{23}$Na($\alpha,n)^{26}$Al reaction together with the $^{23}$Na($\alpha,p)^{26}$Mg reaction. Fig. \ref{fig:an_total} gives a comparison of the total cross section of the $^{23}$Na($\alpha,n)^{26}$Al reaction obtained in this work and  previous measurements by \citet{Norman82}, \citet{Skelton87} and statistical model calculations performed by \citet{Mohr15}. While the cross sections of Refs. \cite{Norman82} and \cite{Skelton87} are taken in small energy steps (for $E_{c.m.}<$4 MeV), we have verified that the average of their data over the energy width of an individual strips of the MUSIC detector agrees with our measurement.  Therefore, a good agreement is obtained between the measurements of this study using the MUSIC detector and previous experiments, as well as with statistical model predictions. Although we have not calculated the contribution from the ground state of the $^{23}$Na($\alpha,n)^{26}$Al reaction as it was done in Refs. \cite{Norman82} and \cite{Skelton87}, our work confirm the total cross section obtained in their work. 

\section{ Summary}

We have performed a simultaneous measurement of excitation functions of the $^{23}$Na$(\alpha,p)^{26}$Mg and $^{23}$Na$(\alpha,n)^{26}$Al reactions, which are important for the understanding of $^{26}$Al production in massive stars. The experiment was carried out making use of a MUlti-Sampling Ionization Chamber and the advantages of inverse kinematics, which gives $\sim$100\% detection efficiency of the reaction products. The cross sections measured for these reactions were found to be in good agreement with previous measurements and statistical model calculations for both, the $^{23}$Na$(\alpha,p)^{26}$Mg and $^{23}$Na$(\alpha,n)^{26}$Al reactions. Furthermore, the cross section of the $^{23}$Na$(\alpha,p)^{26}$Mg reaction was extended to higher energies. Discrepancies of the cross sections for the $^{23}$Na$(\alpha,p)^{26}$Mg reactions in previous measurements have now been settled. 

In addition, we have presented a technique to measure simultaneously excitation functions of angle and excitation-energy integrated cross sections of $(\alpha,p)$ and $(\alpha,n)$ reactions. Because the beam particles and the reaction products are measured in the same detector, problems with normalization of the cross section are avoided. This is an efficient way to study astrophysically important reactions because a large range of the excitation function for two different reactions is covered with a single beam energy.

\begin{acknowledgments}
The authors are grateful to Dr. P. Mohr for helpful discussions. This material is based upon work supported by the U.S. Department of Energy, Office of Science, Office of Nuclear Physics, 
under contract number DE-AC02-06CH11357. The authors J. L. and D. S. G. also acknowledge the support by the U.S. Department of Energy, Office of Science, Office of Nuclear Science, under Award No. DE-FG02-96ER40978.
This research used resources of ANL's ATLAS facility, which is DOE Office of Science
User Facility.
\end{acknowledgments}

\bibliography{myrefs}

\begin{thebibliography}{18}%
\makeatletter
\providecommand \@ifxundefined [1]{%
 \@ifx{#1\undefined}
}%
\providecommand \@ifnum [1]{%
 \ifnum #1\expandafter \@firstoftwo
 \else \expandafter \@secondoftwo
 \fi
}%
\providecommand \@ifx [1]{%
 \ifx #1\expandafter \@firstoftwo
 \else \expandafter \@secondoftwo
 \fi
}%
\providecommand \natexlab [1]{#1}%
\providecommand \enquote  [1]{``#1''}%
\providecommand \bibnamefont  [1]{#1}%
\providecommand \bibfnamefont [1]{#1}%
\providecommand \citenamefont [1]{#1}%
\providecommand \href@noop [0]{\@secondoftwo}%
\providecommand \href [0]{\begingroup \@sanitize@url \@href}%
\providecommand \@href[1]{\@@startlink{#1}\@@href}%
\providecommand \@@href[1]{\endgroup#1\@@endlink}%
\providecommand \@sanitize@url [0]{\catcode `\\12\catcode `\$12\catcode
  `\&12\catcode `\#12\catcode `\^12\catcode `\_12\catcode `\%12\relax}%
\providecommand \@@startlink[1]{}%
\providecommand \@@endlink[0]{}%
\providecommand \url  [0]{\begingroup\@sanitize@url \@url }%
\providecommand \@url [1]{\endgroup\@href {#1}{\urlprefix }}%
\providecommand \urlprefix  [0]{URL }%
\providecommand \Eprint [0]{\href }%
\providecommand \doibase [0]{http://dx.doi.org/}%
\providecommand \selectlanguage [0]{\@gobble}%
\providecommand \bibinfo  [0]{\@secondoftwo}%
\providecommand \bibfield  [0]{\@secondoftwo}%
\providecommand \translation [1]{[#1]}%
\providecommand \BibitemOpen [0]{}%
\providecommand \bibitemStop [0]{}%
\providecommand \bibitemNoStop [0]{.\EOS\space}%
\providecommand \EOS [0]{\spacefactor3000\relax}%
\providecommand \BibitemShut  [1]{\csname bibitem#1\endcsname}%
\let\auto@bib@innerbib\@empty
\bibitem [{\citenamefont {Diehl}\ \emph {et~al.}(1995)\citenamefont {Diehl},
  \citenamefont {Dupraz}, \citenamefont {Bennett}, \citenamefont {Bloemen},
  \citenamefont {Hermsen}, \citenamefont {Kn\"odlseder}, \citenamefont
  {Lichti}, \citenamefont {Sch\"onfelder}, \citenamefont {Steinle},
  \citenamefont {Strong}, \citenamefont {Swanenburg}, \citenamefont
  {Varendorff},\ and\ \citenamefont {Winkler}}]{Diehl95}%
  \BibitemOpen
  \bibfield  {author} {\bibinfo {author} {\bibfnamefont {R.}~\bibnamefont
  {Diehl}}, \bibinfo {author} {\bibfnamefont {C.}~\bibnamefont {Dupraz}},
  \bibinfo {author} {\bibfnamefont {K.}~\bibnamefont {Bennett}}, \bibinfo
  {author} {\bibfnamefont {H.}~\bibnamefont {Bloemen}}, \bibinfo {author}
  {\bibfnamefont {W.}~\bibnamefont {Hermsen}}, \bibinfo {author} {\bibfnamefont
  {J.}~\bibnamefont {Kn\"odlseder}}, \bibinfo {author} {\bibfnamefont
  {G.}~\bibnamefont {Lichti}}, \bibinfo {author} {\bibfnamefont
  {V.}~\bibnamefont {Sch\"onfelder}}, \bibinfo {author} {\bibfnamefont
  {H.}~\bibnamefont {Steinle}}, \bibinfo {author} {\bibfnamefont
  {A.}~\bibnamefont {Strong}}, \bibinfo {author} {\bibfnamefont
  {B.}~\bibnamefont {Swanenburg}}, \bibinfo {author} {\bibfnamefont
  {M.}~\bibnamefont {Varendorff}}, \ and\ \bibinfo {author} {\bibfnamefont
  {C.}~\bibnamefont {Winkler}},\ }\href
  {{http://adsabs.harvard.edu/abs/1995A%26A...298..445D}} {\bibfield  {journal}
  {\bibinfo  {journal} {Astronomy \& Astrophysics}\ }\textbf {\bibinfo {volume}
  {298}},\ \bibinfo {pages} {445} (\bibinfo {year} {1995})}\BibitemShut
  {NoStop}%
\bibitem [{\citenamefont {Diehl}\ \emph {et~al.}(2006)\citenamefont {Diehl},
  \citenamefont {Halloin}, \citenamefont {Kretschmer}, \citenamefont {Lichti},
  \citenamefont {Sch\"onfelder}, \citenamefont {Strong}, \citenamefont {von
  Kienlin}, \citenamefont {Wang}, \citenamefont {Jean}, \citenamefont
  {Kn\"odlseder}, \citenamefont {Roques}, \citenamefont {Weidenspointner},
  \citenamefont {Schanne}, \citenamefont {Hartmann}, \citenamefont {Winkler},\
  and\ \citenamefont {Wunderer}}]{Diehl06}%
  \BibitemOpen
  \bibfield  {author} {\bibinfo {author} {\bibfnamefont {R.}~\bibnamefont
  {Diehl}}, \bibinfo {author} {\bibfnamefont {H.}~\bibnamefont {Halloin}},
  \bibinfo {author} {\bibfnamefont {K.}~\bibnamefont {Kretschmer}}, \bibinfo
  {author} {\bibfnamefont {G.~G.}\ \bibnamefont {Lichti}}, \bibinfo {author}
  {\bibfnamefont {V.}~\bibnamefont {Sch\"onfelder}}, \bibinfo {author}
  {\bibfnamefont {A.~W.}\ \bibnamefont {Strong}}, \bibinfo {author}
  {\bibfnamefont {A.}~\bibnamefont {von Kienlin}}, \bibinfo {author}
  {\bibfnamefont {W.}~\bibnamefont {Wang}}, \bibinfo {author} {\bibfnamefont
  {P.}~\bibnamefont {Jean}}, \bibinfo {author} {\bibfnamefont {J.}~\bibnamefont
  {Kn\"odlseder}}, \bibinfo {author} {\bibfnamefont {J.-P.}\ \bibnamefont
  {Roques}}, \bibinfo {author} {\bibfnamefont {G.}~\bibnamefont
  {Weidenspointner}}, \bibinfo {author} {\bibfnamefont {S.}~\bibnamefont
  {Schanne}}, \bibinfo {author} {\bibfnamefont {D.~H.}\ \bibnamefont
  {Hartmann}}, \bibinfo {author} {\bibfnamefont {C.}~\bibnamefont {Winkler}}, \
  and\ \bibinfo {author} {\bibfnamefont {C.}~\bibnamefont {Wunderer}},\ }\href
  {\doibase 10.1038/nature04364} {\bibfield  {journal} {\bibinfo  {journal}
  {Nature}\ }\textbf {\bibinfo {volume} {439}},\ \bibinfo {pages} {45}
  (\bibinfo {year} {2006})}\BibitemShut {NoStop}%
\bibitem [{\citenamefont {Iliadis}\ \emph {et~al.}(2011)\citenamefont
  {Iliadis}, \citenamefont {Champagne}, \citenamefont {Chieffi},\ and\
  \citenamefont {Limongi}}]{Illiadis11}%
  \BibitemOpen
  \bibfield  {author} {\bibinfo {author} {\bibfnamefont {C.}~\bibnamefont
  {Iliadis}}, \bibinfo {author} {\bibfnamefont {A.}~\bibnamefont {Champagne}},
  \bibinfo {author} {\bibfnamefont {A.}~\bibnamefont {Chieffi}}, \ and\
  \bibinfo {author} {\bibfnamefont {M.}~\bibnamefont {Limongi}},\ }\href
  {http://stacks.iop.org/0067-0049/193/i=1/a=16} {\bibfield  {journal}
  {\bibinfo  {journal} {ApJS}\ }\textbf {\bibinfo {volume} {193}},\ \bibinfo
  {pages} {16} (\bibinfo {year} {2011})}\BibitemShut {NoStop}%
\bibitem [{\citenamefont {Kuperus}(1964)}]{Kuperus64}%
  \BibitemOpen
  \bibfield  {author} {\bibinfo {author} {\bibfnamefont {J.}~\bibnamefont
  {Kuperus}},\ }\href {\doibase doi:10.1016/0031-8914(64)90052-7} {\bibfield
  {journal} {\bibinfo  {journal} {Physica (Amsterdam)}\ }\textbf {\bibinfo
  {volume} {30}},\ \bibinfo {pages} {2253} (\bibinfo {year}
  {1964})}\BibitemShut {NoStop}%
\bibitem [{\citenamefont {Whitmire}\ and\ \citenamefont
  {Davids}(1974)}]{Whitmire74}%
  \BibitemOpen
  \bibfield  {author} {\bibinfo {author} {\bibfnamefont {D.~P.}\ \bibnamefont
  {Whitmire}}\ and\ \bibinfo {author} {\bibfnamefont {C.~N.}\ \bibnamefont
  {Davids}},\ }\href {\doibase 10.1103/PhysRevC.9.996} {\bibfield  {journal}
  {\bibinfo  {journal} {Phys. Rev. C}\ }\textbf {\bibinfo {volume} {9}},\
  \bibinfo {pages} {996} (\bibinfo {year} {1974})}\BibitemShut {NoStop}%
\bibitem [{\citenamefont {Almaraz-Calderon}\ \emph {et~al.}(2014)\citenamefont
  {Almaraz-Calderon}, \citenamefont {Bertone}, \citenamefont {Alcorta},
  \citenamefont {Albers}, \citenamefont {Deibel}, \citenamefont {Hoffman},
  \citenamefont {Jiang}, \citenamefont {Marley}, \citenamefont {Rehm},\ and\
  \citenamefont {Ugalde}}]{Almaraz14}%
  \BibitemOpen
  \bibfield  {author} {\bibinfo {author} {\bibfnamefont {S.}~\bibnamefont
  {Almaraz-Calderon}}, \bibinfo {author} {\bibfnamefont {P.~F.}\ \bibnamefont
  {Bertone}}, \bibinfo {author} {\bibfnamefont {M.}~\bibnamefont {Alcorta}},
  \bibinfo {author} {\bibfnamefont {M.}~\bibnamefont {Albers}}, \bibinfo
  {author} {\bibfnamefont {C.~M.}\ \bibnamefont {Deibel}}, \bibinfo {author}
  {\bibfnamefont {C.~R.}\ \bibnamefont {Hoffman}}, \bibinfo {author}
  {\bibfnamefont {C.~L.}\ \bibnamefont {Jiang}}, \bibinfo {author}
  {\bibfnamefont {S.~T.}\ \bibnamefont {Marley}}, \bibinfo {author}
  {\bibfnamefont {K.~E.}\ \bibnamefont {Rehm}}, \ and\ \bibinfo {author}
  {\bibfnamefont {C.}~\bibnamefont {Ugalde}},\ }\href {\doibase
  10.1103/PhysRevLett.112.152701} {\bibfield  {journal} {\bibinfo  {journal}
  {Phys. Rev. Lett.}\ }\textbf {\bibinfo {volume} {112}},\ \bibinfo {pages}
  {152701} (\bibinfo {year} {2014})}\BibitemShut {NoStop}%
\bibitem [{\citenamefont {Tomlinson}\ \emph {et~al.}(2015)\citenamefont
  {Tomlinson}, \citenamefont {Fallis}, \citenamefont {Laird}, \citenamefont
  {Fox}, \citenamefont {Akers}, \citenamefont {Alcorta}, \citenamefont
  {Bentley}, \citenamefont {Christian}, \citenamefont {Davids}, \citenamefont
  {Davinson}, \citenamefont {Fulton}, \citenamefont {Galinski}, \citenamefont
  {Rojas}, \citenamefont {Ruiz}, \citenamefont {de~S\'er\'eville},
  \citenamefont {Shen},\ and\ \citenamefont {Shotter}}]{Tomlinson15}%
  \BibitemOpen
  \bibfield  {author} {\bibinfo {author} {\bibfnamefont {J.~R.}\ \bibnamefont
  {Tomlinson}}, \bibinfo {author} {\bibfnamefont {J.}~\bibnamefont {Fallis}},
  \bibinfo {author} {\bibfnamefont {A.~M.}\ \bibnamefont {Laird}}, \bibinfo
  {author} {\bibfnamefont {S.~P.}\ \bibnamefont {Fox}}, \bibinfo {author}
  {\bibfnamefont {C.}~\bibnamefont {Akers}}, \bibinfo {author} {\bibfnamefont
  {M.}~\bibnamefont {Alcorta}}, \bibinfo {author} {\bibfnamefont {M.~A.}\
  \bibnamefont {Bentley}}, \bibinfo {author} {\bibfnamefont {G.}~\bibnamefont
  {Christian}}, \bibinfo {author} {\bibfnamefont {B.}~\bibnamefont {Davids}},
  \bibinfo {author} {\bibfnamefont {T.}~\bibnamefont {Davinson}}, \bibinfo
  {author} {\bibfnamefont {B.~R.}\ \bibnamefont {Fulton}}, \bibinfo {author}
  {\bibfnamefont {N.}~\bibnamefont {Galinski}}, \bibinfo {author}
  {\bibfnamefont {A.}~\bibnamefont {Rojas}}, \bibinfo {author} {\bibfnamefont
  {C.}~\bibnamefont {Ruiz}}, \bibinfo {author} {\bibfnamefont {N.}~\bibnamefont
  {de~S\'er\'eville}}, \bibinfo {author} {\bibfnamefont {M.}~\bibnamefont
  {Shen}}, \ and\ \bibinfo {author} {\bibfnamefont {A.~C.}\ \bibnamefont
  {Shotter}},\ }\href {\doibase 10.1103/PhysRevLett.115.052702} {\bibfield
  {journal} {\bibinfo  {journal} {Phys. Rev. Lett.}\ }\textbf {\bibinfo
  {volume} {115}},\ \bibinfo {pages} {052702} (\bibinfo {year}
  {2015})}\BibitemShut {NoStop}%
\bibitem [{\citenamefont {Howard}\ \emph {et~al.}(2015)\citenamefont {Howard},
  \citenamefont {Munch}, \citenamefont {Fynbo}, \citenamefont {Kirsebom},
  \citenamefont {Laursen}, \citenamefont {Diget},\ and\ \citenamefont
  {Hubbard}}]{Howard15}%
  \BibitemOpen
  \bibfield  {author} {\bibinfo {author} {\bibfnamefont {A.~M.}\ \bibnamefont
  {Howard}}, \bibinfo {author} {\bibfnamefont {M.}~\bibnamefont {Munch}},
  \bibinfo {author} {\bibfnamefont {H.~O.~U.}\ \bibnamefont {Fynbo}}, \bibinfo
  {author} {\bibfnamefont {O.~S.}\ \bibnamefont {Kirsebom}}, \bibinfo {author}
  {\bibfnamefont {K.~L.}\ \bibnamefont {Laursen}}, \bibinfo {author}
  {\bibfnamefont {C.~A.}\ \bibnamefont {Diget}}, \ and\ \bibinfo {author}
  {\bibfnamefont {N.~J.}\ \bibnamefont {Hubbard}},\ }\href {\doibase
  10.1103/PhysRevLett.115.052701} {\bibfield  {journal} {\bibinfo  {journal}
  {Phys. Rev. Lett.}\ }\textbf {\bibinfo {volume} {115}},\ \bibinfo {pages}
  {052701} (\bibinfo {year} {2015})}\BibitemShut {NoStop}%
\bibitem [{\citenamefont {Almaraz-Calderon}\ \emph {et~al.}(2015)\citenamefont
  {Almaraz-Calderon}, \citenamefont {Bertone}, \citenamefont {Alcorta},
  \citenamefont {Albers}, \citenamefont {Deibel}, \citenamefont {Hoffman},
  \citenamefont {Jiang}, \citenamefont {Marley}, \citenamefont {Rehm},\ and\
  \citenamefont {Ugalde}}]{Almaraz15}%
  \BibitemOpen
  \bibfield  {author} {\bibinfo {author} {\bibfnamefont {S.}~\bibnamefont
  {Almaraz-Calderon}}, \bibinfo {author} {\bibfnamefont {P.~F.}\ \bibnamefont
  {Bertone}}, \bibinfo {author} {\bibfnamefont {M.}~\bibnamefont {Alcorta}},
  \bibinfo {author} {\bibfnamefont {M.}~\bibnamefont {Albers}}, \bibinfo
  {author} {\bibfnamefont {C.~M.}\ \bibnamefont {Deibel}}, \bibinfo {author}
  {\bibfnamefont {C.~R.}\ \bibnamefont {Hoffman}}, \bibinfo {author}
  {\bibfnamefont {C.~L.}\ \bibnamefont {Jiang}}, \bibinfo {author}
  {\bibfnamefont {S.~T.}\ \bibnamefont {Marley}}, \bibinfo {author}
  {\bibfnamefont {K.~E.}\ \bibnamefont {Rehm}}, \ and\ \bibinfo {author}
  {\bibfnamefont {C.}~\bibnamefont {Ugalde}},\ }\href {\doibase
  10.1103/PhysRevLett.115.179901} {\bibfield  {journal} {\bibinfo  {journal}
  {Phys. Rev. Lett.}\ }\textbf {\bibinfo {volume} {115}},\ \bibinfo {pages}
  {179901} (\bibinfo {year} {2015})}\BibitemShut {NoStop}%
\bibitem [{\citenamefont {Norman}\ \emph {et~al.}(1982)\citenamefont {Norman},
  \citenamefont {Timothy}, \citenamefont {Lesko},\ and\ \citenamefont
  {Peter}}]{Norman82}%
  \BibitemOpen
  \bibfield  {author} {\bibinfo {author} {\bibfnamefont {E.~B.}\ \bibnamefont
  {Norman}}, \bibinfo {author} {\bibfnamefont {E.~C.}\ \bibnamefont {Timothy}},
  \bibinfo {author} {\bibfnamefont {K.~T.}\ \bibnamefont {Lesko}}, \ and\
  \bibinfo {author} {\bibfnamefont {S.}~\bibnamefont {Peter}},\ }\href
  {\doibase doi:10.1016/0375-9474(82)90283-4} {\bibfield  {journal} {\bibinfo
  {journal} {Nucl. Phys. A}\ }\textbf {\bibinfo {volume} {390}},\ \bibinfo
  {pages} {561} (\bibinfo {year} {1982})}\BibitemShut {NoStop}%
\bibitem [{\citenamefont {Skelton}\ \emph {et~al.}(1987)\citenamefont
  {Skelton}, \citenamefont {Kavanagh},\ and\ \citenamefont
  {Sargood}}]{Skelton87}%
  \BibitemOpen
  \bibfield  {author} {\bibinfo {author} {\bibfnamefont {R.~T.}\ \bibnamefont
  {Skelton}}, \bibinfo {author} {\bibfnamefont {R.~W.}\ \bibnamefont
  {Kavanagh}}, \ and\ \bibinfo {author} {\bibfnamefont {D.~G.}\ \bibnamefont
  {Sargood}},\ }\href {\doibase 10.1103/PhysRevC.35.45} {\bibfield  {journal}
  {\bibinfo  {journal} {Phys. Rev. C}\ }\textbf {\bibinfo {volume} {35}},\
  \bibinfo {pages} {45} (\bibinfo {year} {1987})}\BibitemShut {NoStop}%
\bibitem [{\citenamefont {Carnelli}\ \emph {et~al.}(2015)\citenamefont
  {Carnelli}, \citenamefont {Almaraz-Calderon}, \citenamefont {Rehm},
  \citenamefont {Albers}, \citenamefont {Alcorta}, \citenamefont {Bertone},
  \citenamefont {Digiovine}, \citenamefont {Esbensen}, \citenamefont
  {Fernández~Niello}, \citenamefont {Henderson}, \citenamefont {Jiang},
  \citenamefont {Lai}, \citenamefont {Marley}, \citenamefont {Nusair},
  \citenamefont {Palchan-Hazan}, \citenamefont {Pardo}, \citenamefont {Paul},\
  and\ \citenamefont {Ugalde}}]{Carnelli15}%
  \BibitemOpen
  \bibfield  {author} {\bibinfo {author} {\bibfnamefont {P.~F.~F.}\
  \bibnamefont {Carnelli}}, \bibinfo {author} {\bibfnamefont {S.}~\bibnamefont
  {Almaraz-Calderon}}, \bibinfo {author} {\bibfnamefont {K.~E.}\ \bibnamefont
  {Rehm}}, \bibinfo {author} {\bibfnamefont {M.}~\bibnamefont {Albers}},
  \bibinfo {author} {\bibfnamefont {M.}~\bibnamefont {Alcorta}}, \bibinfo
  {author} {\bibfnamefont {P.~F.}\ \bibnamefont {Bertone}}, \bibinfo {author}
  {\bibfnamefont {B.}~\bibnamefont {Digiovine}}, \bibinfo {author}
  {\bibfnamefont {H.}~\bibnamefont {Esbensen}}, \bibinfo {author}
  {\bibfnamefont {J.}~\bibnamefont {Fernández~Niello}}, \bibinfo {author}
  {\bibfnamefont {D.}~\bibnamefont {Henderson}}, \bibinfo {author}
  {\bibfnamefont {C.~L.}\ \bibnamefont {Jiang}}, \bibinfo {author}
  {\bibfnamefont {J.}~\bibnamefont {Lai}}, \bibinfo {author} {\bibfnamefont
  {S.~T.}\ \bibnamefont {Marley}}, \bibinfo {author} {\bibfnamefont
  {O.}~\bibnamefont {Nusair}}, \bibinfo {author} {\bibfnamefont
  {T.}~\bibnamefont {Palchan-Hazan}}, \bibinfo {author} {\bibfnamefont {R.~C.}\
  \bibnamefont {Pardo}}, \bibinfo {author} {\bibfnamefont {M.}~\bibnamefont
  {Paul}}, \ and\ \bibinfo {author} {\bibfnamefont {C.}~\bibnamefont
  {Ugalde}},\ }\href {\doibase 10.1016/j.nima.2015.07.030} {\bibfield
  {journal} {\bibinfo  {journal} {Nucl. Inst. and Meth. in Phys. Res. Section
  A}\ }\textbf {\bibinfo {volume} {799}},\ \bibinfo {pages} {197} (\bibinfo
  {year} {2015})}\BibitemShut {NoStop}%
\bibitem [{\citenamefont {Bair}\ and\ \citenamefont {Haas}(1973)}]{Bair73}%
  \BibitemOpen
  \bibfield  {author} {\bibinfo {author} {\bibfnamefont {J.~K.}\ \bibnamefont
  {Bair}}\ and\ \bibinfo {author} {\bibfnamefont {F.~X.}\ \bibnamefont
  {Haas}},\ }\href {\doibase 10.1103/PhysRevC.7.1356} {\bibfield  {journal}
  {\bibinfo  {journal} {Phys. Rev. C}\ }\textbf {\bibinfo {volume} {7}},\
  \bibinfo {pages} {1356} (\bibinfo {year} {1973})}\BibitemShut {NoStop}%
\bibitem [{\citenamefont {Avila}\ \emph {et~al.}()\citenamefont {Avila},
  \citenamefont {Rehm}, \citenamefont {Almaraz-Calderon}, \citenamefont
  {Ayangeakaa}, \citenamefont {Dickerson}, \citenamefont {Hoffman},
  \citenamefont {Jiang}, \citenamefont {Kay}, \citenamefont {Lai},
  \citenamefont {Nusair}, \citenamefont {Pardo}, \citenamefont
  {Santiago-Gonzalez}, \citenamefont {Talwar},\ and\ \citenamefont
  {Ugalde}}]{AvilaNIM16}%
  \BibitemOpen
  \bibfield  {author} {\bibinfo {author} {\bibfnamefont {M.~L.}\ \bibnamefont
  {Avila}}, \bibinfo {author} {\bibfnamefont {K.~E.}\ \bibnamefont {Rehm}},
  \bibinfo {author} {\bibfnamefont {S.}~\bibnamefont {Almaraz-Calderon}},
  \bibinfo {author} {\bibfnamefont {A.~D.}\ \bibnamefont {Ayangeakaa}},
  \bibinfo {author} {\bibfnamefont {C.}~\bibnamefont {Dickerson}}, \bibinfo
  {author} {\bibfnamefont {C.~R.}\ \bibnamefont {Hoffman}}, \bibinfo {author}
  {\bibfnamefont {C.~L.}\ \bibnamefont {Jiang}}, \bibinfo {author}
  {\bibfnamefont {B.~P.}\ \bibnamefont {Kay}}, \bibinfo {author} {\bibfnamefont
  {J.}~\bibnamefont {Lai}}, \bibinfo {author} {\bibfnamefont {O.}~\bibnamefont
  {Nusair}}, \bibinfo {author} {\bibfnamefont {R.~C.}\ \bibnamefont {Pardo}},
  \bibinfo {author} {\bibfnamefont {D.}~\bibnamefont {Santiago-Gonzalez}},
  \bibinfo {author} {\bibfnamefont {R.}~\bibnamefont {Talwar}}, \ and\ \bibinfo
  {author} {\bibfnamefont {C.}~\bibnamefont {Ugalde}},\ }\href@noop {}
  {\bibinfo  {journal} {Nucl. Inst. and Meth. in Phys. Res. Section A (To be
  published)}\ }\BibitemShut {NoStop}%
\bibitem [{\citenamefont {Kubik}\ \emph {et~al.}(1987)\citenamefont {Kubik},
  \citenamefont {Elmore}, \citenamefont {Hemmick}, \citenamefont {Gove},
  \citenamefont {Fehn}, \citenamefont {Teng}, \citenamefont {Jiang},\ and\
  \citenamefont {Tullai}}]{Kubik87}%
  \BibitemOpen
\bibfield  {journal} {  }\bibfield  {author} {\bibinfo {author} {\bibfnamefont
  {P.~W.}\ \bibnamefont {Kubik}}, \bibinfo {author} {\bibfnamefont
  {D.}~\bibnamefont {Elmore}}, \bibinfo {author} {\bibfnamefont {T.~K.}\
  \bibnamefont {Hemmick}}, \bibinfo {author} {\bibfnamefont {H.~E.}\
  \bibnamefont {Gove}}, \bibinfo {author} {\bibfnamefont {U.}~\bibnamefont
  {Fehn}}, \bibinfo {author} {\bibfnamefont {R.~T.}\ \bibnamefont {Teng}},
  \bibinfo {author} {\bibfnamefont {S.}~\bibnamefont {Jiang}}, \ and\ \bibinfo
  {author} {\bibfnamefont {S.}~\bibnamefont {Tullai}},\ }\href {\doibase
  doi:10.1016/0168-583X(87)90222-9} {\bibfield  {journal} {\bibinfo  {journal}
  {Nucl. Inst. and Meth. in Phys. Res. Section B}\ }\textbf {\bibinfo {volume}
  {29}},\ \bibinfo {pages} {138} (\bibinfo {year} {1987})}\BibitemShut
  {NoStop}%
\bibitem [{\citenamefont {Mohr}(2015)}]{Mohr15}%
  \BibitemOpen
  \bibfield  {author} {\bibinfo {author} {\bibfnamefont {P.}~\bibnamefont
  {Mohr}},\ }\href {\doibase 10.1140/epja/i2015-15056-5} {\bibfield  {journal}
  {\bibinfo  {journal} {EpJ A}\ }\textbf {\bibinfo {volume} {51}},\ \bibinfo
  {pages} {1} (\bibinfo {year} {2015})},\ \bibinfo {note} {and private
  communication}\BibitemShut {NoStop}%
\bibitem [{\citenamefont {Ziegler}\ \emph {et~al.}(2010)\citenamefont
  {Ziegler}, \citenamefont {Ziegler},\ and\ \citenamefont {P.}}]{SRIM}%
  \BibitemOpen
  \bibfield  {author} {\bibinfo {author} {\bibfnamefont {J.~F.}\ \bibnamefont
  {Ziegler}}, \bibinfo {author} {\bibfnamefont {M.~D.}\ \bibnamefont
  {Ziegler}}, \ and\ \bibinfo {author} {\bibfnamefont {B.~J.}\ \bibnamefont
  {P.}},\ }\href {\doibase doi:10.1016/j.nimb.2010.02.091} {\bibfield
  {journal} {\bibinfo  {journal} {Nucl. Inst. and Meth. in Phys. Res. Section
  B}\ }\textbf {\bibinfo {volume} {268}},\ \bibinfo {pages} {1818} (\bibinfo
  {year} {2010})}\BibitemShut {NoStop}%
\bibitem [{\citenamefont {Tarsov}\ and\ \citenamefont {Bazin}(2008)}]{LISE}%
  \BibitemOpen
  \bibfield  {author} {\bibinfo {author} {\bibfnamefont {O.~B.}\ \bibnamefont
  {Tarsov}}\ and\ \bibinfo {author} {\bibfnamefont {D.}~\bibnamefont {Bazin}},\
  }\href {\doibase doi:10.1016/j.nimb.2008.05.110} {\bibfield  {journal}
  {\bibinfo  {journal} {Nucl. Inst. and Meth. in Phys. Res. Section B}\
  }\textbf {\bibinfo {volume} {266}},\ \bibinfo {pages} {4657} (\bibinfo {year}
  {2008})}\BibitemShut {NoStop}%
\end{thebibliography}%

\end{document}